\documentstyle[preprint,aps,eqsecnum]{revtex}
 
\def\beq#1{\begin{equation} \label{#1}}
\def\eeq{\end{equation}}
\def\bra#1{\left\langle #1\right\vert}
\def\ket#1{\left\vert #1\right\rangle}
\def\epsp{\epsilon^{\prime}}

\def\NPB{{ Nucl. Phys.} B}
\def\PLB{{ Phys. Lett.} B}

\def\PRD{{ Phys. Rev.} D}
\begin{document}
{
\tighten
\preprint {\vbox{
 \hbox{WIS-96/27/Jun-PH}
 \hbox{TAUP 2346-96}
 \hbox{hep-ph/9607201} 
 \hbox{}
}}

\title{Flavor Oscillations from a Spatially Localized Source \\
             A Simple General Treatment}
 
\author{Yuval Grossman\,$^a$ and Harry J. Lipkin\,$^{a,b}$}
 
\address{ \vbox{\vskip 0.truecm}
  $^a\;$Department of Particle Physics \\
  Weizmann Institute of Science, Rehovot 76100, Israel \\
\vbox{\vskip 0.truecm}
$^b\;$School of Physics and Astronomy \\
Raymond and Beverly Sackler Faculty of Exact Sciences \\
Tel Aviv University, Tel Aviv, Israel}
 
\maketitle
 
\begin{abstract}%
A unique description avoiding confusion is presented for all flavor oscillation
experiments in which particles of a definite flavor are emitted from a
localized source. The probability for finding a particle with the wrong flavor
must vanish at the position of the source for all times. This condition
requires flavor--time and flavor--energy factorizations which determine uniquely
the flavor mixture observed at a detector in the oscillation region; i.e.
where the overlaps between the wave packets for different mass eigenstates
are almost complete. Oscillation periods calculated for ``gedanken''
time-measurement experiments are shown to give the correct measured
oscillation wave length in space when multiplied by the group velocity.
Examples of neutrinos propagation in a weak field and in a gravitational field
are given. In these cases the relative phase is modified differently for
measurements in space and time. Energy-momentum (frequency-wave number)
and space-time descriptions are complementary, equally valid and
give the same results. The two identical phase shifts obtained describe the
same physics; adding them together to get a factor of two is double counting.
 
\end{abstract}%
 
} 
 
\newpage

\section {Introduction}
Flavor oscillations are observed when a source creates a particle
which is a mixture of two or more mass eigenstates, and
a different mixture is observed in a detector. Such oscillations have been
observed in the neutral kaon and B--meson systems.
In neutrino experiments
it is still unclear whether the eigenstates indeed have
different masses and whether oscillations can be observed.
Considerable confusion has arisen in the description of such
experiments in quantum mechanics \cite{Kayser,NeutHJL},
with questions arising about time dependence and production
reactions \cite{GoldS}, and defining precisely what 
is observed in an experiment \cite{Pnonexp}.
Many calculations describe ``gedanken" experiments and require some
recipe for applying the results
to a real experiment \cite{MMNIETO}.
 
We resolve this confusion by noting
and applying one simple general feature of all practical experiments.
The size of the source is small in comparison with the
oscillation wave length to be measured, and
a unique well--defined flavor mixture is emitted by the source; e.g. electron
neutrinos in
a neutrino oscillation experiment. The particles emitted from the source must
therefore be described by a wave packet which satisfies a simple general
boundary condition: the probability amplitude for finding a particle having
the wrong flavor at the source must vanish at all times.
 
This boundary condition requires factorization of the flavor and time dependence
at the position of the source. Since the energy dependence is the Fourier
transform of the time dependence, this factorization also
implies that the flavor dependence of the wave packet is independent of energy
at the position of the source.
In a realistic oscillation experiment the relative phase is important when the
oscillation length is of the same order as the distance between the
source and the detector.
In that case this
flavor--energy factorization holds over the entire distance between the source
and detector. The boundary condition then determines
the relative phase of components in the wave function with
different mass having the same energy and different momenta. Thus
any flavor oscillations observed as a function of the distance between the
source and the detector are
described by considering only the interference between
a given set of states having the same energy. All questions of coherence,
relative phases of components in the wave function with different energies and
possible entanglements with other degrees of freedom are thus avoided.
 
Many formulations describe flavor oscillations in time produced by
interference between states with equal momenta and different energies.
These ``gedanken" experiments
have flavor oscillations in time over all space including the source.
We show rigorously that the ratio of
the wave length of the real spatial oscillation to the period of
the gedanken time oscillation is
just the group velocity of the wave packet.
 
\section{Universal Boundary Condition}
 
We now show how the results of a flavor oscillation experiment are completely
determined by the propagation dynamics and the boundary condition that the
probability of observing a particle of the wrong flavor at the position of the
source at any time must vanish. We choose
for example a neutrino oscillation experiment with a source of neutrinos of a
given flavor, say electron neutrinos\footnote{%
For simplicity, we do not consider possible effects of physics beyond the
Standard Model on neutrino interactions\cite{Yuval}.
The generalization to this case is straightforward.}.
The dimensions of the source are
sufficiently small in comparison with the distance to the detector so that it
can be considered a point source at the origin.
The neutrino wave function for this experiment may be a very
complicated wave packet, but a sufficient condition for our analysis is to
require it to describe a pure
$\nu_e$ source at $x = 0$; i.e. the probability of finding a
$\nu_\mu$ or $\nu_\tau$ at $x = 0$ is zero.
 
We first consider propagation in
free space, where the masses and momenta $p_i$ satisfy the usual condition
\beq{WW1b}
p_i^2 = E^2 - m_i^2\,.
\eeq
We expand the neutrino wave function in energy eigenstates
\beq{WW1a}
\psi = \int g(E) dE e^{-iEt}\cdot \sum_{i=1}^3 c_i e^{i p_i\cdot x}
\ket {\nu_i}\,,
\eeq
where $\ket {\nu_i}$ denote the three neutrino mass eigenstates and the
coefficients $c_i$ are energy-independent.
Each energy eigenstate has three terms, one for each mass eigenstate.
In order to avoid spurious flavor oscillations at the source the
particular linear combination of these three terms
required to describe this experiment must be
a pure $\nu_e$ state at $x = 0$ for each individual energy component.
Thus the coefficients $c_i$ satisfy the conditions
\beq{WW2}
\sum_{i=1}^3 c_i \bra {\nu_i}\nu_\mu \rangle =
\sum_{i=1}^3 c_i \bra {\nu_i}\nu_\tau \rangle = 0\,.
\eeq
The momentum of each of the
three components is determined by the energy and the neutrino masses. The
propagation of this energy eigenstate, the relative phases of its three mass
components  and its flavor mixture at the detector are completely determined by
the energy-momentum kinematics for the three mass eigenstates.
 
The exact form of the energy wave packet described by the function $g(E)$ is
irrelevant at this stage. The components with different energies may be coherent
or incoherent, and they may be ``entangled" with other degrees of freedom of
the system. For the case where a neutrino is produced together with
an electron in a weak decay the function $g(E)$ can also be a function
$g(\vec p_e,E)$ of the electron momentum as well as the neutrino energy.
The neutrino degrees of freedom observed at the detector will then be described
by a density matrix after the electron degrees of freedom have been properly
integrated out, taking into account any measurements on the electron. However,
none of these considerations can introduce a neutrino of the wrong flavor at the
position of the source.
 
Since the momenta $p_i$ are
energy-dependent the factorization does not hold at finite distance.
At very large values of $x$ the wave packet must separate into individual
wave packets with different masses traveling with different velocities
\cite{Nus,Kayser}.
However, for the conditions of a realistic oscillation experiment this
separation has barely begun and the overlap of the wave packets with different
masses is essentially 100\%. Under these conditions the flavor--energy
factorization introduced at the source is still an excellent approximation at
the detector.
 
The flavor mixture at the detector given by substituting the
detector coordinate into Eq. (\ref{WW1a})
can be shown to be the same for all the
energy eigenstates except for completely negligible small differences.
For example, for the case of two neutrinos with energy $E$ and mass eigenstates
$m_1$ and $m_2$ the relative phase of the two neutrino waves at a distance $x$
is:
\beq{WW3a}
\delta \phi(x)= (p_1 - p_2)\cdot x =
{{(p_1^2 - p_2^2)}\over{(p_1 + p_2)}}\cdot x  =
{{\Delta m^2}\over{(p_1 + p_2)}}\cdot x\,,
\eeq
where $\Delta m^2 \equiv m_2^2-m_1^2$.
Since the neutrino mass difference is very small compared to all neutrino
momenta and energies, we use
$|m_2 - m_1| \ll p \equiv (1/2)(p_1 + p_2)$.
Thus we can rewrite Eq. (\ref{WW3a})
keeping terms only of first order in $\Delta m^2$
\beq{WW4a}
\delta \phi(x) =
{{\Delta m^2}\over{2p}}\cdot x =
- \left({{\partial p}\over{\partial (m^2)}}\right)_{E}
\Delta m^2\cdot x\,,
\eeq
where the standard relativistic energy-momentum relation (\ref{WW1b}) gives
the change in energy or momentum with mass when the other is fixed,
\beq{WW4b}
\left({{2E\partial E}\over{\partial (m^2)}}\right)_{p}
= - \left({{2p\partial p}\over{\partial (m^2)}}\right)_{E} = 1\,.
\eeq

Thus we have a complete
solution to the oscillation problem and can give the neutrino flavor as a
function of the distance to the detector by examining the behavior of a single
energy eigenstate. The flavor--energy factorization enables the result to be
obtained without considering any interference effects between different energy
eigenstates. The only information needed
to predict the neutrino oscillations is the behavior
of a linear combination of the three mass eigenstates having
the same energy and different momenta.
All effects of interference or relative phase between
components of the wave function with different energies are time dependent
and are required to vanish at the source, where the flavor is time independent.
This time independence also holds at the detector as long as
there is significant overlap
between the wave packets for different mass states.
The conditions for the validity of this overlap condition are discussed below.
 
Neutrino states with the same energy and different momenta are
relevant rather than vice versa
because the measurement is in space, not time, and flavor--time factorization
holds in a definite region in space.
 
\section{Relation between Real and Gedanken Experiments}
We now derive the relation between our result (\ref{WW3a}) which comes from
interference between states with the same energy and different momenta and
the standard treatments using states with the same momentum and different
energies \cite{booknu}.
For the case of two neutrinos with momentum $p$ and mass eigenstates
$m_1$ and $m_2$ the relative phase of the two neutrino waves at a time $t$ is:
\beq{WW6}
\delta \phi(t)= (E_2 - E_1)\cdot t =
\left({{\partial E}\over{\partial (m^2)}}\right)_{p}
\Delta m^2 \cdot t =
- \left({{\partial p}\over{\partial (m^2)}}\right)_{E}
\Delta m^2 \cdot {{p}\over{E}} \cdot t\,,
\eeq
where we have substituted Eq. (\ref{WW4b}).
This is equal to the result (\ref{WW4a})
if we make the commonly used substitution
\beq{WW7b}
x = {{p}\over{E}} \cdot t = v t\,.
\eeq
 
This is now easily generalized to include cases where external fields can
modify the relation (\ref{WW1b}),
but where the mass eigenstates are not mixed. The
extension to propagation in a medium which mixes mass eigenstates e.g. by the
MSW effect \cite{revMSW} is in principle the same, but
more complicated in practice and not considered here.
The relation between energy, momentum and mass is described by an
arbitrary dispersion relation
\beq{WW15}
f(E, p, m^2) = 0\,,
\eeq
where the function $f$ can also be a slowly varying function of the distance
$x$. In that case, the momentum $p$ for fixed $E$ is also a slowly varying
function of $x$. We take this into account by expressing Eq. (\ref{WW4a})
as a differential equation, and defining
the velocity $v$ by the conventional expression for the group
velocity,
\beq{WW16a}
{{\partial^2  \phi(x)}\over{\partial x \partial (m^2)}}=
-\left({{\partial p}\over{\partial (m^2)}}\right)_{E}
= {{1}\over{v}}\cdot \left({{\partial E}\over{\partial (m^2)}}\right)_{p}
\;, \qquad
v \equiv \left({{\partial E}\over{\partial p}}\right)_{(m^2)}\,.
\eeq
Treatments describing real experiments measuring
distances and ``gedanken" experiments measuring time
are seen to be rigorously equivalent if the
group velocity (\ref{WW16a}) relates the two results.
Note that the group velocity and not
the phase velocity enters into this relation.
The relations (\ref{WW16a}) are trivial and obvious
for the case of neutrinos propagating in free space, and gives
Eq. (\ref{WW7b}).
However, it becomes nontrivial for more complicated cases. Two such
cases are presented in the following.

\section{Description in Terms of Time Behavior}
 
The specific form of the wave packet given by the function $g(E)$ in
Eq. (\ref{WW1a}) describes the
Fourier transform of the time behavior as seen at
$ x = 0 $. This time behavior changes as the packet moves from
source to detector. Components corresponding to different
mass eigenstates move with different velocities. When the centers of the wave
packets have moved a distance $x_c$ they have separated by a distance
\beq{WW5a}
\delta x_c = {{\delta v}\over{v}}\cdot x  \approx
 {{\delta p}\over{p}}\cdot x  =
{{\Delta m^2 }\over{2p^2}}\cdot x \,,\qquad
\delta v \equiv v_1 - v_2 \,,\qquad \delta p \equiv p_1 - p_2\,,
\eeq
where
$v_1$, $v_2$ and $v$  denote the individual group velocities of the two wave
packets and an average group velocity,
and we have assumed that
$m_i^2 = E_i^2 - p_i^2 \ll p_i^2$.
This separation between the wave packet centers
produces a phase displacement between the waves at the
detector, $\delta\phi(x) = p\, \delta x_c$, which is seen to give exactly the
same phase shift as Eq. (\ref{WW3a}).
The group velocity which determines the separation between the wave packets is
relevant and not the phase velocity.
 
Further insight into the relation between different treatments
is seen by rewriting the phase shift Eq. (\ref{WW3a})
in terms of the distance $\xi \equiv x - x_c $ between the point $x$ and the
center of the wave packet as the sum of
the relative phase shift between the centers of the two wave packets
$\delta \phi(x_c)$ at a fixed time and a ``correction" to this phase shift
because the centers of the wave packets arrive at the detector at different
times. To first order in the small quantities  $ \delta x $ and $\delta p$
\beq{WW75a}
\delta x_c + \delta \xi = 0  \,,\qquad  \delta \phi(x) = \delta (xp) =
x \delta p +  p \delta x_c + p \delta \xi
= \delta \phi(x_c) + p \delta \xi\,,
\eeq
\beq{WW75b}
\delta \phi(x_c) \equiv x \delta p + p \delta x_c =
{{\Delta m^2 }\over{p}}\cdot x \,,\qquad
p \delta \xi = - p \delta x_c = - {{\Delta m^2 }\over{2p}}\cdot x\,.
\eeq
Writing the phase shift in this form and neglecting the
``correction" leads to an overestimate of the phase by a factor of two, while
adding the ``correction" to the correct interpretation (\ref{WW6})
of the gedanken experiment can lead to double counting.
 
We see here simply another description of the
same physics used in the derivation of Eq. (\ref{WW3a}),
using the complementarity of energy-momentum and space-time
formulations. They are two ways of getting the same answer, not two different
effects that must be added.
 
The same complementarity is seen in the interference between two classical
wave packets moving with slightly different velocities. Even without using
the quantum mechanical relations with energy and momentum there are two possible
descriptions, one using space and time variables and one using frequency and
wave length. The two descriptions are Fourier transforms of one another and
give the same result. Adding the two results is double counting.
 
We now apply this picture of two wave packets traveling with
slightly different velocities to examine
the time-dependent probability amplitude for a neutrino wave
seen at the detector when it is emitted from the source in a flavor
eigenstate denoted by $\ket{f_1}$.
The $x$ dependences of the amplitude and other parameters are suppressed since
we only need their values at the position of the detector.
\beq{YY807}
\ket{\Psi(t)} = e^{i \phi_o(t)}
\left[ \cos \theta A(t)  \ket{m_1} +
\sin \theta A(t + \tau) e^{i  \phi(\tau) }\ket{m_2}\right]\,,
\eeq
where $\ket{m_1}$ and $\ket{m_2}$ denote the two mass eigenstates and
$\theta$ is a mixing angle defining the flavor eigenstates denoted by
$\ket{f_1}$ and $\ket{f_2}$ in terms of the mass eigenstates,
\beq{YY802}
\ket{f_1} = \cos \theta \ket{m_1} + \sin \theta \ket{m_2} \,,\qquad
\ket{f_2} = \sin \theta \ket{m_1} - \cos  \theta \ket{m_2}\,,
\eeq
\beq{YY804}
\tau(x) = {{x}\over{v_2}} - {{x}\over{v_1}} \approx {\delta v \over v^2} x
\approx {{\Delta m^2 }\over{2p^2 \,v}} x\,,
\eeq
where $\delta v /v$ is always defined for components in the different
mass eigenstates having the same energy and the small variation in
$\delta v / v$ over the wave packet is neglected.
We express each mass eigenstate wave function as the product of a magnitude
$A(x)$ and a phase. The universal boundary condition requires $A$ to be the
same for both mass eigenstates at the source. The wave functions spread
with distance and may become much broader at the detector. However
the difference in shape between the two mass eigenstates is shown below
to be negligible at the detector under experimental conditions
where oscillations are observable. Their center
difference is described by the time
displacement $\tau$.
 
The probability amplitudes for observing the flavor eigenstates at the detector
are
\beq{YY808}
\langle f_1 \ket{\Psi(t)} = e^{i \phi_o(t)}
\left[\cos^2 \theta A(t) e^{i  \phi(\tau) }+
\sin^2 \theta A(t+ \tau)  \right]\,,
\eeq
\beq{YY809}
\langle f_2 \ket{\Psi(t)} = e^{i \phi_o(t)} \sin \theta \cos \theta
\left[ A(t) e^{i  \phi(\tau) } -
A(t+ \tau)  \right]\,.
\eeq
The relative probabilities that flavors
$f_1$ and $f_2$ are observed at the detector are
\beq{YY8010}
P(f_1,\tau) = \int dt |\langle f_1 \ket{\Psi(t)}|^2 =
 1 - {\sin^2 (2\theta) \over 2} \Big[1 - O(\tau) \cos \phi(\tau)\Big]\,,
\eeq
\beq{YY8011}
P(f_2,\tau) = \int dt |\langle f_2 \ket{\Psi(t)}|^2 =
 {\sin^2 (2\theta) \over 2} \Big[1 - O(\tau) \cos \phi(\tau) \Big]\,,
\eeq
where the amplitude normalization and the overlap function $O(\tau)$ are given
by
\beq{YY8013}
\int dt |A(t)|^2 = 1 \,,\qquad
O(\tau) \equiv \int dt A(t + \tau) A(t)\,.
\eeq
When the overlap is complete, $O(\tau) \approx 1$,
the results (\ref{YY8010}) and (\ref{YY8011}) reduce
to the known result obtained by assuming plane waves \cite{booknu} and using
\beq{taurel}
\phi(\tau)=p \delta x_c = p\,v\, \tau \approx {{\Delta m^2 }\over{2p}}x\,.
\eeq
An explicit example for the calculation of the overlap function can be found
in Ref. \cite{GKL} where the shape function $A$ was taken to be a Gaussian.
 
We now examine the spreading of the wave functions while traveling from the
source to the detector. The length of the wave packet in space
$L_w(0)$ in
the vicinity of the source
must be sufficiently large to contain a large number $N_w$ of wave lengths
$\lambdabar$ in order to define a phase. This then determines the spread of
the momentum, $\delta p_w$, and velocity, $\delta v_w$, in the wave packet
\beq{YY601}
L_w(0) = N_w \lambdabar = {{N_w}\over{p}} \;,\qquad
{{\delta p_w}\over{p}}= {{\delta v_w}\over{v}} = {{1}\over{N_w}}\,.
\eeq
The spreading of the wave packet in traveling from the source to
the point $x$ is
\beq{YY602}
{{L_w(x) - L_w(0)}\over{L_w(0)}}  =
{{\delta v_w}\over{v}}\cdot {{x}\over{ L_w(0)}}=
{{\delta p_w}\over{p}}\cdot {{x \cdot p}\over{ N_w}}=
{{x \cdot p}\over{ N_w^2}}\,.
\eeq
The difference in the spreading of the wave packets for the different mass
eigenstates is then seen to be negligible for distances $x$ where the
oscillation phase shift $\delta \phi(x)$ is of order unity
\beq{YY603}
{{\partial}\over{\partial (m^2)}}\left({{L_w(x) - L_w(0)}\over{L_w(0)}}\right)
\cdot \Delta m^2 = {{\partial p}\over{\partial (m^2)}}\cdot \Delta m^2 \cdot
{{x }\over{ N_w^2}} = {{\delta \phi(x) }\over{ N_w^2}}\,.
\eeq
 
The different mass eigenstates separate as a result of the velocity differences.
Eventually the wave packet separates into distinct packets,
one for each mass, moving with different velocities. The separation destroys
the flavor--energy and flavor--time factorizations and introduces a time
dependence in the flavor observable in principle at a given large distance.
In practice the detailed time dependence is not measurable and only the
attenuation of the oscillation expressed by the overlap function $O(\tau)$
is seen. When the wave
packets for different masses no longer overlap there is no longer any
coherence and there are no further oscillations \cite{Nus}.
The result (\ref{WW3a}) applies for the case where the separation
(\ref{WW5a}) is small compared to the length in space of the wave
packet; i.e. when the eventual separation of the wave packets has barely begun
and can be neglected.

\section{Fuzziness in time}
 
The oscillations can be described either in space or in time.
But the distance between the source and the detector is known in a realistic
experiment to much higher accuracy then the time interval.
Thus the interval between the two events of creation and detection has a sharp
distance and a fuzzy time in the laboratory system.
A Lorentz transformation to a different frame necessarily mixes distance and
time and makes both fuzzy in a complicated manner. For this reason one must
be careful in interpreting any results obtained in other frames than the
laboratory system. The proper time interval between the two events is always
fuzzy.
 
The fuzziness of the time is an essential feature of the experiment since the
wave packet has a finite length $L_w$ in space. The probability of
observing the particle at the detector is spread over the time interval
\beq{YY611}
2 \delta t \equiv  {{L_w}\over{v}} = {{L_w E}\over{p}}\,.
\eeq
The proper time interval
$\tau$ between emission and detection is given by
\beq{YY612}
\tau^2 = (t \pm \delta t)^2 - x^2 = x^2 \left[ {{m^2}\over{p^2}} +
{{L_w^2 E^2}\over{4x^2 p^2}}  \pm {{L_w E^2}\over{x p^2}}\right]
= x^2 \, {{m^2}\over{p^2}} \left[ 1 + {{E^2}\over{m^2}}\cdot \left(
{{L_w^2 }\over{4x^2 }}  \pm {{L_w }\over{x }}\right)\right]\,.
\eeq
This uncertainty in the proper time interval due to the finite length of the
wave packet cannot be neglected.

The waves describing the propagation of different mass eigenstates can be
coherent at the detector only if the overlap function $O(\tau)$ given
by Eq. (\ref{YY8013}) is nearly unity. Thus
the time interval between creation and detection is not
precisely determined and subject to quantum--mechanical fluctuations.
The length $L_w$ of the wave packet created at the source must be sufficiently
long to prevent the determination of its velocity by a
time measurement with the precision needed to identify the mass eigenstate.
 
The small dimensions of the source introduce a momentum uncertainty essential
for the coherence of the waves of different mass eigenstates. The wave packet
describing the experiment must necessarily contain components from different
mass eigenstates with the same energy and different momenta.
 
Conventional experiments
measure distances to a precision with an error tiny in comparison with
the oscillation wave length to be measured. This is easily achieved in the
laboratory. In a ``gedanken" experiment where oscillations in time are
measured, the experimental apparatus must
measure times to a precision with an error tiny in comparison with
the oscillation period to be measured. One might envision an experiment
which measures the time the oscillating particle is created
by observing another particle emitted at the same time; e.g. an
electron emitted in a beta decay together with the neutrino whose oscillation
is observed. But if both the time and position of the created particle are
measured with sufficient precision a very sharp wave packet is created and the
mass eigenstates moving with different velocities quickly separate, the
overlap function $O(\tau)$ approaches zero and there is no coherence and no
oscillation.
 
In reality, when both $x$ and $t$ are measured there are fluctuations in their
values. Using $v=x/t$ the fluctuations in $x$ and $t$ must be large enough to
make the velocity fuzzy. 
Then, in order to have oscillation we need
the fuzziness in velocity to be much larger than the difference between the two
group velocities,
$ \delta v_w \gg \delta v$.
This is the case in a real experiment.
Typical values are
\cite{exper}
$E = O(10\; MeV)$; $x=O(10^2\;m)$; $t=O(10^{-6}\; sec)$
and the relevant masses that can be probed are $\Delta m^2 = O(1\; eV^2)$.
Then, $\delta v = O(10^{-12})$.
Since $\delta v_w \approx dx/x + dt/t$ we
see that the accuracies needed to measure the separate velocities are
$dx = O(10^{-10}\;m)$ and $dt = O(10^{-18}\; sec)$,
far from the ability of present technology.
This calculation can also be performed for all terrestrial experiments, finding
that the present technology is not yet sufficiently precise to destroy coherence
and prevent oscillations from being observed.
\section{Examples}
 
The relations (\ref{WW16a}) are trivial and obvious
for the case of neutrinos propagating in free space. However, it becomes
nontrivial for more complicated cases.
In this section we present two nontrivial examples:
Neutrino in a (flavor blind) weak field and
neutrino in a gravitational field.
These are only examples, in real life the effects we discuss tend to be
very small, and consequently negligible. Yet, these examples demonstrate
how to get the phase shift, and how to move from the description
in terms of time to that of space using the group velocity.
 
In these examples we calculate the phase difference for a known
beam with known energy. We consider a source and a detector in vacuum
and investigate the effect of inserting a field (either weak or gravitational)
between them.
 
\subsection{Neutrino in a weak field}
We consider neutrino travel in a flavor--blind medium.
The medium changes the dispersion relation \cite{revMSW} by introducing
the potential $V$ describing the scattering in the medium
\beq{we1}
(E+V)^2-p^2=m^2\,.
\eeq
For simplicity we assume that $V$ is independent of $x$ but
can depend upon $E$. The phase difference in space and in time
are then given by
\beq{we2}
\delta \phi(x)=
- \left({{\partial p}\over{\partial (m^2)}}\right)_{E}
\Delta m^2\cdot x  =
 { \Delta m^2 \over 2 p} \cdot x  \approx
 { \Delta m^2 \over 2 p_o} \, (1-\epsilon) \cdot x\,,
\eeq
\beq{we3}
\delta \phi(t)=
- \left({{\partial E}\over{\partial (m^2)}}\right)_{p}
\Delta m^2\cdot t =
{ \Delta m^2 \over 2 (E+ V)(1+{d V \over d E})} \cdot t
\approx   { \Delta m^2 \over 2 E} \, {1-\epsilon \over 1+ \epsp}  \cdot t\,,
\eeq
where $p \approx E+V$ and $p_o  \approx E$ are the momentum in
the medium and in free space, respectively. We work to first order in
$\epsilon$ and $\epsp$ defined as
\beq{defeps}
\epsilon \equiv {V\over E}\,, \qquad \epsp={d V\over d E}\,.
\eeq
 
We learn that the medium effect is {\it different} for the two cases
\beq{we4}
{\delta \phi(x) \over \delta \phi_o(x)} = 1-\epsilon \;, \qquad
{\delta \phi(t) \over \delta \phi_o(t)} = {1-\epsilon \over 1+\epsp}\,,
\eeq
where $\delta \phi_o(x)$ and $\delta \phi_o(t)$ denote the values
respectively of $\delta \phi(x)$ and $\delta \phi(t)$ for the case where
$V =0$. To move from one description to the other we need the
group velocity
\beq{we5}
v = \left({{\partial E}\over{\partial p}}\right)_{(m^2)}
= {p \over (E + V)(1+{d V\over d  E})} = {1 \over 1+\epsp}\,.
\eeq
Using $t \rightarrow x/v = x(1+\epsp)$ in (\ref{we3}) we get (\ref{we2}).
We see that by using the correct velocity one can relate the two descriptions
and the results are the same.
 
Note that our example is not realistic.
In the Standard Model the neutral current interactions (that are flavor blind)
are energy independent. Then, $\epsp=0$ and
the group velocity is not changed from its vacuum value.
 
This example has a simple optical analog. Consider an optical interference
experiment (e.g. a two slit experiment) with a glass inserted in the
light path. A measurement in space will gain a larger phase shift due to
the travel in the medium. The light travels slower in the medium and when it
reaches the detector the optical path is longer.
 
\subsection{Neutrino in a gravitation field}
We consider neutrino travel in a gravitational field.
This has recently been treated in Refs. \cite{AB,comAB,ABnew}.
We compare two cases:
one when the neutrino travel is in free space, a second when a gravitational
field is inserted in the path. We assume that the gravitational field is
sufficiently small to leave
the (Newtonian) distance unaffected by its insertion.
One example is the possible effect of the moon on solar neutrinos
when the moon is close to solar eclipse. Then we shall see that
the gravitational field of the moon affects the phase.
 
We assume: (1) The semi--classical limit;
(2) The weak field limit;
(3) Nearly Newtonian gravitational fields.
The first assumption \cite{Stod} says that
gravity is not quantized and its effect is introduced by
a nonflat space-time metric
$g_{\mu \nu} \ne \eta_{\mu \nu}$,
where $\eta_{\mu \nu}={\rm diag}(1,-1,-1,-1)$ is the flat metric.
The second assumption \cite{book}
says that we can use the linear
approximation. Then, gravity is treated as an external field on a
flat space time and we expand
\beq{gr2}
g_{\mu \nu} = \eta_{\mu \nu} + h_{\mu \nu}\,,
\eeq
with $|h_{\mu \nu}| \ll 1$.
The third assumption  \cite{book}
says that the gravitational field originates from a massive static source.
Then
\beq{defne}
h_{\mu \mu} = 2 \Phi(\vec{x})\,,
\qquad  h_{\mu \nu} = 0 ~~ {\rm for} ~~ \mu \ne \nu \,,
\eeq
where $\Phi(\vec{x})$ is the Newtonian potential
(e.g. $\Phi(\vec{x})=-G\,M/|\vec{x}|$
for a spherically symmetric object with mass $M$).
We emphasize that $h_{00} = h_{ii}$ but $\eta_{00} = - \eta_{ii}$.
This sign difference turns out to be important.
 
The dispersion relation in a curved space-time is \cite{book}
\beq{(yy6)}
g_{\mu\nu} p^\mu p^\nu = m^2\,,
\eeq
where $p^\mu =md x^\mu/ds$ is the local momentum, and
$ds$ is the distance element of general relativity:
$ds^2=g_{\mu\nu} dx^\mu dx^\nu$.
We consider neutrinos that travel in space-time from $A$ to $B$.
The wave function is then \cite{Stod}
\beq{gr4}
\psi=\exp({i \phi}), \qquad \phi = \int_A^B g_{\mu \nu} p^\mu dx^\nu\,.
\eeq
The phase difference in space and in time
are then given by
\beq{yWW4}
\delta \phi(x)=
 \int_A^B g_{11} (p_2-p_1) dx =
\int_A^B \left({g_{11}{\partial p}\over{\partial (m^2)}}\right)_{E}
\Delta m^2 dx\,,
\eeq
\beq{yWW6}
\delta \phi(t)= \int_A^B g_{00} (E_2 - E_1)  dt =
\int_A^B \left({g_{00}{\partial E}\over{\partial (m^2)}}\right)_{p}
\Delta m^2 dt\,.
\eeq
The velocity is then obtained by generalizing
Eq. (\ref{WW16a})
\beq{yWW16b}
v = -\left({g_{00}{\partial E}\over{g_{11} \partial p}}\right)_{(m^2)}\,.
\eeq
Applying this to the dispersion relation we get
\beq{ff1}
\delta \phi(x)=
\int_A^B { \Delta m^2 \over 2 p} dx
\approx \int_A^B (1-\Phi(\vec{x})){ \Delta m^2 \over 2 p_o} dx\,,
\eeq
\beq{ff2}
\delta \phi(t)=
\int_A^B { \Delta m^2 \over 2 E} dt \approx
 \int_A^B (1+\Phi(\vec{x})){ \Delta m^2 \over 2 E_o} dt\,,
\eeq
where $p_o^\mu =md x^\mu/ds_o$ is the usual momentum of
special relativity (global momentum) \cite{Stod}.
We work to first order in $\Phi(\vec{x})$ and we use \cite{book,Stod}
\beq{fcc3}
p \approx p_o (1+\Phi(\vec{x}))\;, \qquad
E \approx E_o (1-\Phi(\vec{x}))\,.
\eeq
Our result (\ref{ff1}) is the one obtained in \cite{AB}.
 
We learn that the gravitational effect is {\it different} for the two cases
\beq{fcc4}
{\delta \phi(x) \over \delta \phi_o(x)} =
{\lambda_o \over \lambda} = 1-\epsilon \;, \qquad
{\delta \phi(t) \over \delta \phi_o(t)} =
{\tau_o \over \tau} =
1+\epsilon\,,
\eeq
where $\lambda$ and $\lambda_o$ denote the wave length of the oscillation in
space for the case with and without the gravitational field respectively
and similarly $\tau$ and $\tau_o$ denote the period of the
oscillation in time for the two cases and we define
\beq{fcc5}
\epsilon \equiv  \int_A^B \Phi(\vec{x}) { \Delta m^2 \over 2 p_o} dx \approx
\int_A^B \Phi(\vec{x}(t)) { \Delta m^2 \over 2 E_o} dt\,.
\eeq
Note that the effect of the gravitational field on the oscillation
wave length $\lambda$ in space is exactly opposite to the effect on
the oscillation period $\tau$ in time.
In order to move from one description to the other we need the
velocity. From (\ref{yWW16b}) we get
\beq{fcc6}
v = {p \over E} \approx 1+2\Phi(\vec{x})\,,
\eeq
which is the known result of the
speed of light in a gravitational field \cite{book}.
Using $t \rightarrow x/v \approx  x(1-2\Phi(\vec{x}))$
in (\ref{ff2}) we get (\ref{ff1}).
 
It is important to understand the meaning of this shift.
We work in the example given before, and examine the effect of
moon gravity on solar neutrinos. Since we assume that
the earth--sun distance is not changed the effect can be viewed in two
equivalent ways. One is the point of view of the linearized
theory of gravity \cite{book}.
Then, space-time is flat and gravity is treated as
a tensor field. In this approach, taken by \cite{AB},
the neutrino travels the same distance with and without the moon, but
gravity slows down the neutrino, thus it has a longer ``optical'' path
and a larger phase is acquired. The second point of view is to work within
the framework of general relativity. Then gravity is treated by changing the
metric into curved space-time. In this approach, taken by \cite{comAB},
the neutrino always travels in free space. However,
when the moon comes close to the sun-earth line
the distance the neutrino has to travel is larger.
The effect of gravity is then moved into the boundary of the integral,
and we see that a larger phase is acquired.
Of course, if one compares two experimental setups with and without
gravity with the same curved distance in both cases
there is no effect \cite{comAB}.
 
The analog of the two points of view is the famous ``bending of light''.
When light travels near the sun it is bent. This
can be understood in two equivalent ways. Either that gravity acts on the light
and curves its path, or that the space near the sun is curved.
With either point of view, the final result is the same,
we observe the bending  of the light.
 
It is instructive to see how the effect can be obtained from the description in
terms of time behavior. Then we just need the distance between the centers
of the wave packets (\ref{WW5a}), or
equivalently, the time between their arrivals.
This time difference can be calculated by taking two classical relativistic
particles with the same energy and different masses leaving the source.
Then, the time difference of their arrival can be calculated.
The result shows the gravitational effect. The time delay is sensitive to
the presence of the gravitational field in the path.
 
Finally, we comment about the interplay between the gravitational and
the MSW effects. In order for the gravitational
effect to be appreciable a very strong gravitational field
must be present. This may be the case in supernova. In this
case there is also a weak field originating from the matter
in the star, or from the neutrinos themselves \cite{revMSW}.
In general, this tends to significantly reduce the mixing angles 
\cite{wol}
very near to the value zero in which the flavor eigenstate
$\nu_e$ is also a mass eigenstate. In the adiabatic
limit a neutrino created in matter
in a mass eigenstate remains a single mass eigenstate throughout its career.
Its flavor can flip in a manner that explains the solar neutrino puzzle
\cite{booknu}, but
there are no oscillations and the gravitational phase cannot be observed.
Of course gravity effects can be important beyond the effect
on the coherent phase. We do not study such effects here.

 
\section{Conclusions}
 
The complete description of a flavor oscillation experiment requires knowledge
of the density matrix for the flavor-mixed state. This depends upon the
production mechanism and possible entanglements with other degrees of freedom
as well as on other dynamical factors which are often ignored. A proton in a
fixed-target experiment is not really free but bound by some kind of effective
potential with characteristic lattice energies like Debye temperatures, which
are of the order of tens of millivolts. This energy scale is no longer
negligible in comparison with mass differences between flavor
eigenstates\cite{Mossb}. The bound proton is not strictly on shell and has
potential as well as kinetic energy. Arguments of Galilean and Lorentz
invariance and separation of center-of-mass motion may not hold for the
kinematics of the production process if the degrees of freedom producing the
binding are neglected.
 
In this paper all these complications are avoided and
a unique prescription has been given for the relative phases of the
contributions from different mass eigenstates to a flavor oscillation
experiment with a localized source having a well defined flavor.
The boundary condition that the probability of observing a particle of the
wrong flavor at the source position must vanish for all times requires a
factorization in flavor and energy of the wave function at the position of the
source. This uniquely determines the wave length of the oscillations observed
at the detector as long as the overlap between wave packets for different
mass eigenstates is maintained at the position of the detector.
 
Whether this wave-packet overlap is sufficiently close to 100\% at the detector
depends upon other parameters in the experiment which determine the detailed
time behavior of the wave packet. If this overlap is appreciable but no longer
nearly complete, the time behavior of the flavor mixture at the
detector can be extremely complicated with leading and trailing edges of the
wave packet being pure mass eigenstates and the intermediate region having a
changing flavor mixture depending upon the relative magnitudes of the
contributing mass eigenstates as well as the relative phases. This detailed
behavior is not observable in practice; only the time integral is measured.
 
A unique prescription has been given for interpreting results of
calculations for ``gedanken" experiments which measure oscillations in time
for components in the
wave packets having the same momentum and different energies. The period
of oscillation in time is related to the wave length of oscillation in space
by the group velocity of the waves.
 
Results are simple in the laboratory system where the positions of the source
and detector are sharp in comparison with all other relevant distances, and
times and proper times must be fuzzy to enable coherent oscillations to be
observed.
 
Two nontrivial examples were given. Neutrinos propagating in weak fields
and in gravitational fields. In both cases the relative phase is modified
by the presence of the field. The phase shift is different for a real
experiment with measurements in space, and for ``gedanken'' experiments done in
time. We show how the group velocity relates the two descriptions.
 
\acknowledgments
We thank Dharam Ahluwalia, Christoph Burgard, Boris Kayser, Pawel Mazur
and Leo Stodolsky for helpful discussions and comments.
One of us (HJL) wishes to thank the Institute of Nuclear Theory at the
University of Washington for its hospitality and to
acknowledge partial support from the U.S. Department of Energy (DOE)
and from the German-Israeli Foundation for Scientific Research and
Development (GIF).
 
{
\tighten

}
 
\end{document}